\documentclass[conference]{IEEEtran}
\IEEEoverridecommandlockouts
\usepackage{cite}
\usepackage{amsmath,amssymb,amsfonts}
\usepackage{algorithmic}
\usepackage[toc, acronym]{glossaries}
\usepackage{graphicx}
\usepackage{multirow}
\usepackage{textcomp}
\usepackage{xcolor}
\usepackage{orcidlink}

\def\BibTeX{{\rm B\kern-.05em{\sc i\kern-.025em b}\kern-.08em
    T\kern-.1667em\lower.7ex\hbox{E}\kern-.125emX}}

\newacronym{dsc}{DSC}{dice similarity coefficient}
\newacronym{gt}{GT}{ground truth}
\newacronym{imc}{IMC}{imaging mass cytometry}
\newacronym{mc}{MC}{missing cells}
\begin{document}

\title{Pushing the limits of cell segmentation models for imaging mass cytometry\\
\thanks{This work was supported by the Engineering and Physical Sciences Research Council [grant number EP/T518177/1].\\
$^{\dagger}$ Contributed equally to the work.\\}
}

\author{\IEEEauthorblockN{Kimberley M. Bird\orcidlink{0009-0003-2395-5803}, 
                          Xujiong Ye\orcidlink{0000-0003-0115-0724},
                          James M. Brown$^{\dagger}$\orcidlink{0000-0001-7636-4554}}
\IEEEauthorblockA{\textit{School of Computer Science} \\
\textit{University of Lincoln}\\
Lincoln, UK}
\and
\IEEEauthorblockN{Alan M. Race$^{\dagger}$\orcidlink{0000-0001-8996-2641}}
\IEEEauthorblockA{\textit{Computer Vision \& AI} \\
\textit{AstraZeneca Computational Pathology GmbH}\\
Munich, Germany}
}

\maketitle

\begin{abstract}
\Acrfull{imc} is a relatively new technique for imaging biological tissue at subcellular resolution. In recent years, learning-based segmentation methods have enabled precise quantification of cell type and morphology, but typically rely on large datasets with fully annotated \acrfull{gt} labels. This paper explores the effects of imperfect labels on learning-based segmentation models and evaluates the generalisability of these models to different tissue types. Our results show that removing 50\% of cell annotations from \acrshort{gt} masks only reduces the \acrfull{dsc} score to 0.874 (from 0.889 achieved by a model trained on fully annotated \acrshort{gt} masks). This implies that annotation time can in fact be reduced by at least half without detrimentally affecting performance. Furthermore, training our single-tissue model on imperfect labels only decreases \acrshort{dsc} by 0.031 on an unseen tissue type compared to its multi-tissue counterpart, with negligible qualitative differences in segmentation. Additionally, bootstrapping the worst-performing model (with 5\% of cell annotations) a total of ten times improves its original \acrshort{dsc} score of 0.720 to 0.829. These findings imply that less time and work can be put into the process of producing comparable segmentation models; this includes eliminating the need for multiple \acrshort{imc} tissue types during training, whilst also providing the potential for models with very few labels to improve on themselves. Source code is available on GitHub: \url{https://github.com/kimberley/ISBI2024}.
\end{abstract}


\section{Introduction}\label{introduction}
\begin{figure}[t]
\centerline{\includegraphics[width=0.5\textwidth]{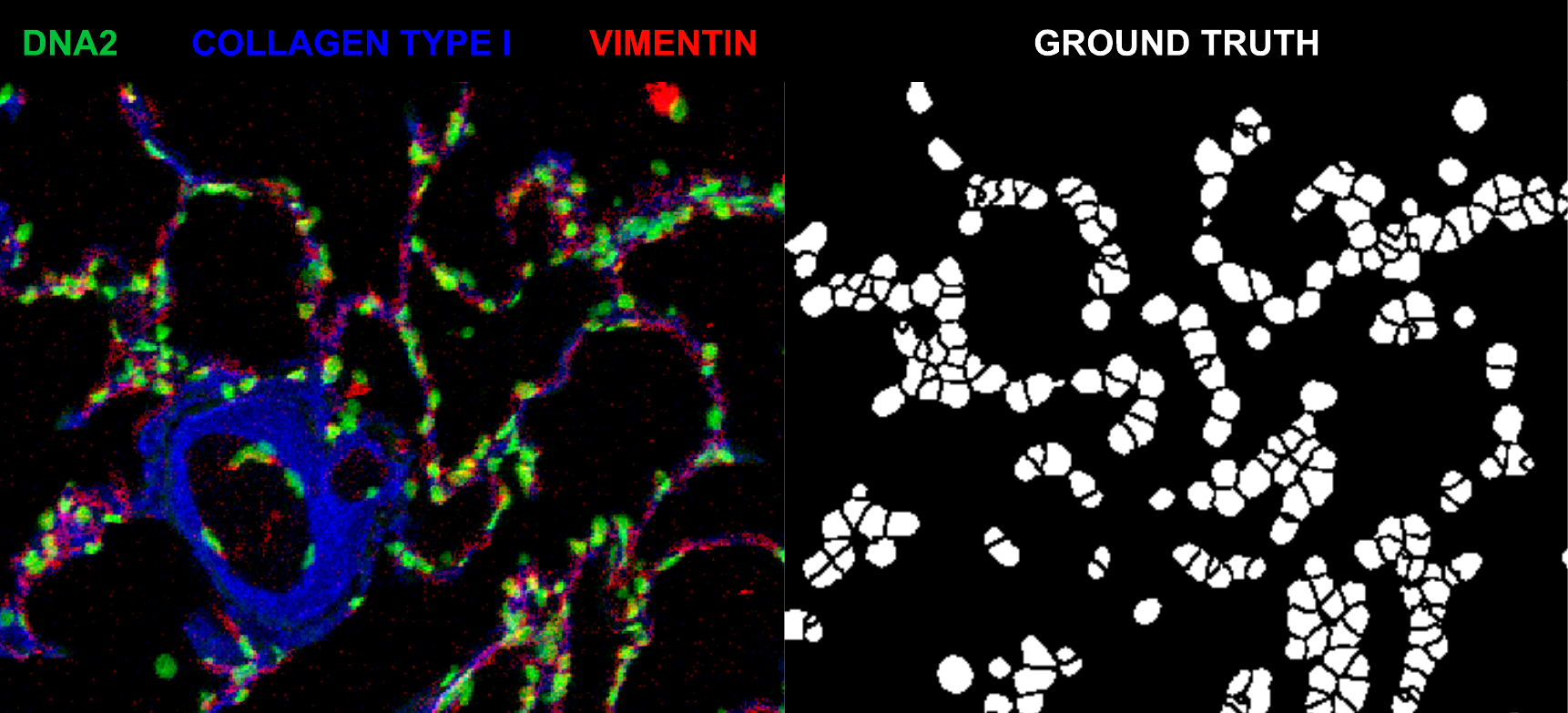}}
\caption{Example \acrshort{imc} image of lung tissue with three biomarkers used throughout this work, next to its corresponding \acrshort{gt} mask. Data from \cite{rendeiro2021}.}
\label{fig-imcchans}
\end{figure}

\Acrfull{imc} is a relatively new technique that uses metal-tagged antibodies to image tissue sections with precise quantification of biological tissue at subcellular resolution (Fig. \ref{fig-imcchans}). These datasets are multi-channel with tens of biomarkers which, in turn, give researchers the ability to visualise several biomarkers simultaneously \cite{giesen2014}. In recent years, \acrshort{imc} has helped to improve our understanding of the tumour microenvironment to improve patient care \cite{glasson2023}, investigating mitochondrial dysfunction \cite{melin2022}, and understanding treatment resistance \cite{zabransky2023}.
Automated and semi-automated segmentation of \acrshort{imc} data allows for precise quantification of cell type and morphology but presents many challenges, particularly with regard to manual annotation of training data. The authors of Mesmer report that it took the equivalent of two years of full-time work to fully annotate TissueNet which comprises over 1 million images \cite{greenwald2022}. As it stands, current learning-based \acrshort{imc} cell segmentation models (such as Mesmer \cite{greenwald2022}, DVP \cite{mund2022}, and MATISSE\cite{baars2021}) typically require large quantities of accurate \acrfull{gt} masks. However, a thorough analysis of label correctness and its effects on model performance across different tissue types has not yet been performed.

In this work, we study the effects of varying levels of label imperfection in \acrshort{imc} data on learning-based cell segmentation models. We simulate the process of generating imperfect labels using classical image processing approaches, and train a U-Net architecture \cite{ronneberger2015} using modified \acrshort{gt} masks with varying levels of label corruption. This includes cases where cells may have been missed by the annotator, along with under- and over-segmentation of cells. We further examine whether imperfectly trained models can generalise to unseen tissue types. Lastly, we perform bootstrapping experiments on models trained with highly corrupted \acrshort{gt}, in order to further push the limits of models trained with minimal cell annotations.

\section{Methods}\label{methods}

\subsection{Data}\label{data}
For the most part of this work, we used an open access lung tissue \acrshort{imc} dataset and the provided \acrshort{gt} masks \cite{rendeiro2021}. It contains a total of 23 samples at a resolution of $1 \mu m$ per pixel from humans with acute lung injury caused by various diseases including COVID-19, as well as healthy subjects (19 with lung disease, 4 without). We utilised channels from six of the 36 biomarkers as detailed in Table \ref{tab-biomarkers}, and used 229 of the 237 images due to mismatched images and masks. Additionally, we used an open access breast tissue \acrshort{imc} dataset (Section \ref{tissue-transfer}) and the provided \acrshort{gt} masks \cite{ali2020}. The acquisitions are based on 483 tumour samples from the METABRIC cohort, totalling 548 images also at a resolution of $1\mu m$ with 37 biomarkers per acquisition, of which only 4 were used in our work (Table \ref{tab-biomarkers}). Both of these datasets' \acrshort{gt} masks were generated using the same methods with cell imaging processing toolkits ilastik \cite{berg2019} and CellProfiler \cite{stirling2021}.

\begin{table}[htbp]
\caption{\acrshort{imc} Biomarkers Used Per Tissue Dataset Per Experiment}
\begin{center}
\renewcommand{\arraystretch}{1.17}
\begin{tabular}{|c|cccc|}
\hline
 & \multicolumn{1}{c|}{\begin{tabular}[c]{@{}c@{}}\textbf{Label}\\ \textbf{Corruption}\end{tabular}} & \multicolumn{1}{c|}{\textbf{Bootstrapping}} & \multicolumn{2}{c|}{\begin{tabular}[c]{@{}c@{}}\textbf{Tissue Type}\\ \textbf{Transferability}\end{tabular}} \\
\hline
\hline
\textbf{Biomarker} & \multicolumn{3}{c|}{\textbf{Lung Tissue}} & \textbf{Breast Tissue} \\
\hline
Histone H3 & \multicolumn{1}{c|}{$\times$} & \multicolumn{1}{c|}{$\times$} & \multicolumn{1}{c|}{$\times$} & $\times$ \\
\hline
Vimentin & \multicolumn{1}{c|}{$\times$} & \multicolumn{1}{c|}{$\times$} & \multicolumn{1}{c|}{$\times$} & $\times$ \\
\hline
Collagen Type I & \multicolumn{1}{c|}{$\times$} & \multicolumn{1}{c|}{$\times$} & \multicolumn{1}{c|}{} & \\
\hline
Vimentin & \multicolumn{1}{c|}{$\times$} & \multicolumn{1}{c|}{$\times$} & \multicolumn{1}{c|}{} & \\
\hline
DNA1 & \multicolumn{1}{c|}{$\times$} & \multicolumn{1}{c|}{$\times$} & \multicolumn{1}{c|}{$\times$} & $\times$ \\
\hline
DNA2 & \multicolumn{1}{c|}{$\times$} & \multicolumn{1}{c|}{$\times$} & \multicolumn{1}{c|}{$\times$} & $\times$ \\
\hline
\end{tabular}
\label{tab-biomarkers}
\end{center}
\end{table}

\subsection{Data Pre-processing}\label{data-preprocessing}
The \acrshort{imc} biomarkers of interest, as detailed in Table \ref{tab-biomarkers}, were extracted from each acquisition for both the lung and breast datasets. The choice of biomarkers was based on the approach previously reported by \cite{rendeiro2021}. Prior to model training, all \acrshort{imc} biomarker channels were normalised using 99th percentile normalisation to mitigate the influence of any extreme pixel values. Percentile normalised images are defined as $\hat{I}=I/P$, where $P$ is set to the 99th percentile based on the image's intensity values. To reduce computational demand, the \acrshort{imc} data and \acrshort{gt} masks were cropped into patches of $128\times128$ pixels with no overlap, adding appropriately sized padding to non-square regions. After model training, the output segmentation masks were unpadded and reconstructed back into their original sizes. We divided each dataset into training, validation, and test splits with a 70:10:20\% ratio, respectively. The lung dataset was stratified by disease state, whereas the breast dataset was randomly divided by acquisition. In total, the lung dataset had 12,929 training, 4,722 validation, and 3,302 testing patches, and the breast dataset had 6,630 training, 1,031 validation, 2,014 testing patches. For the label corruption models, the effect of \acrfull{mc} was simulated by first performing connected component labelling of the \acrshort{gt} masks, and then randomly erasing a fixed percentage of cells. Under- and over-segmentation of cells was simulated using morphological erosion and dilation operations with various kernel sizes.

\subsection{Model Architecture}\label{model-architecture}
The model architecture used in this work is a U-Net with minimal modifications from \cite{ronneberger2015}. The label corruption and bootstrapping models had an input of $128\times128\times6$ using only the lung dataset, and the tissue type transferability models had an input of $128\times128\times4$, using both lung and breast tissue (see Table \ref{tab-biomarkers}). Hyperparameters were optimised for the baseline model and remained unchanged for all other experiments (batch size=64, learning rate=0.01, optimiser=Adam, activation=ReLU, loss=Soft Dice). The final output layer was subjected to sigmoid activation. The U-Net consisted of 5 downsampling steps and 5 upsampling steps, with downsampling filter sizes $=\{32, 64, 128, 256, 512\}$ (reversed for upsampling), and with bottleneck filter size $=512$. Dropout layers were also added throughout the model, with each individual downsampling layer consisting of dropout $=\{0.1, 0.1, 0.2, 0.2, 0.3\}$ respectively (reversed for upsampling), and with bottleneck dropout $=0.4$.

\subsection{Experimental design}\label{experimental-design}
All experiments were developed in Python 3.10.9 and implemented in Keras 2.10.0, using an AMD Ryzen 9 5900X CPU and NVIDIA RTX 3080 10GB GPU.

\paragraph{Label corruption}
The baseline model was trained using the \acrshort{imc} biomarkers as outlined in Table \ref{tab-biomarkers}. A further twelve models were trained with increasing proportions of \acrshort{mc} as follows: 1\%, 2\%, 5\%, 10\%, 20\%, 25\%, 50\%, 75\%, 80\%, 85\%, 90\%, and 95\%. Additionally, four under- and over-segmentation models were trained (based on a model with 50\% of cells removed), with each cell object in the \acrshort{gt} masks randomly eroded or dilated based on a pool of $k\times k$ kernel sizes, $S$, up to a specified maximum size, $k_{max}$, where $S=\{ k \in K : k \leq k_{max} \}$ and $K=\{0, 3, 5, 7, 9\}$ (with $k=0$ having no effect on a cell object).
\paragraph{Tissue type transferability}
Eight models were trained in total, with the first four using only the lung tissue dataset, and the latter four using both the lung and breast tissue datasets. The first four models were trained on the four \acrshort{imc} biomarkers outlined in Table \ref{tab-biomarkers} using the same hyperparameters as the baseline model, based on 10\% \acrshort{mc}, 50\% \acrshort{mc}, and 90\% \acrshort{mc}. The remaining four models were trained in the same way but with both the lung and breast tissue datasets. 
\paragraph{Bootstrapping}
The bootstrapping models were trained based on the model trained with 95\% cells removed from \acrshort{gt} masks. The process consisted of training a model on corrupted \acrshort{gt} data, running inference on the training set from the epoch with the best validation \acrfull{dsc} score, and then updating the \acrshort{gt} masks for the next round of training. We iterated through this process a total of ten times.

\section{Results}\label{results}

\begin{figure}[h]
\centerline{\includegraphics[width=0.5\textwidth]{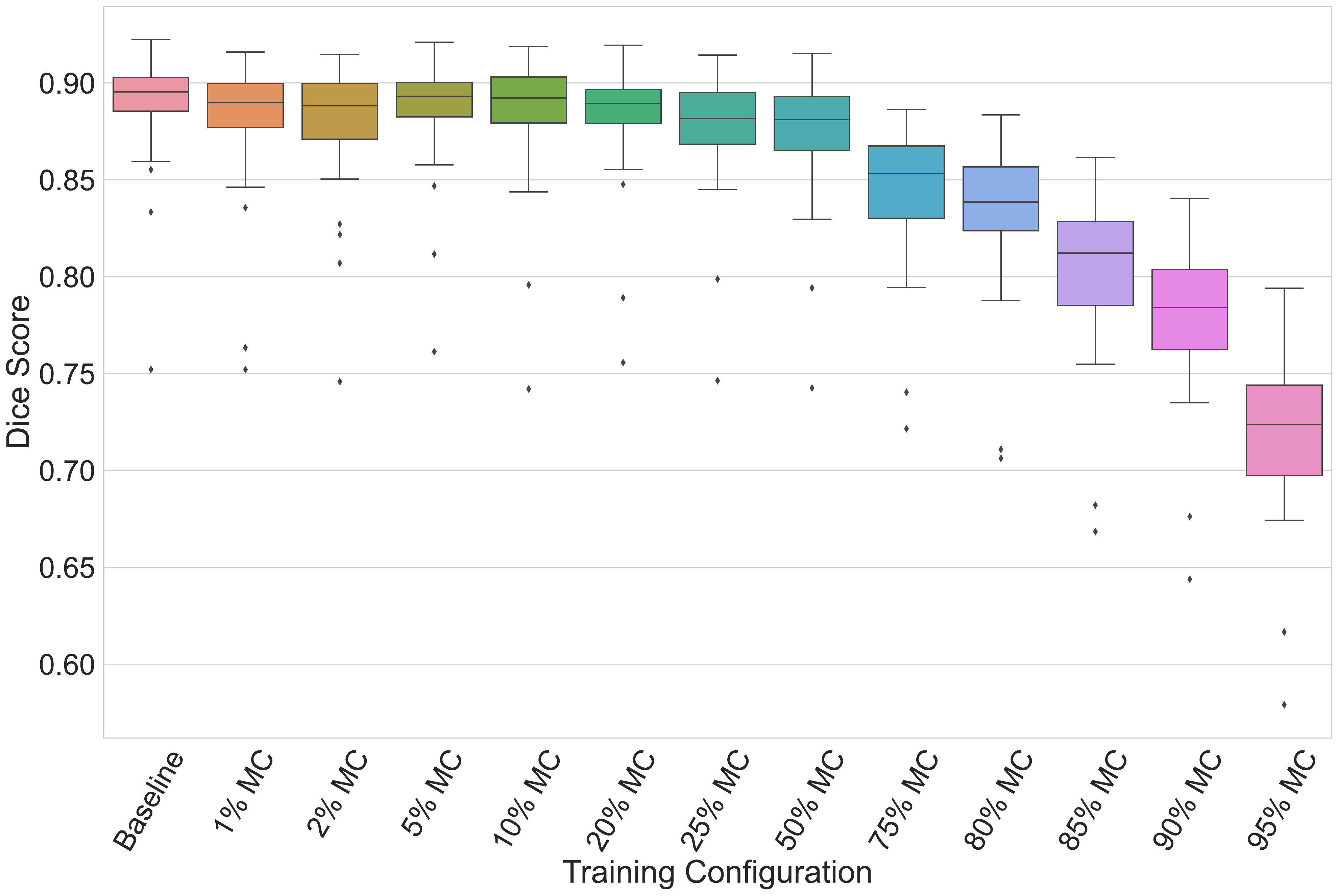}}
\caption{Boxplot comparing segmentation performance with increasing proportions of missing cells from the \acrshort{gt}.}
\label{fig-boxplot-mc}
\end{figure}

\subsection{Label Corruption}\label{label-corruption}
\begin{figure}
\centerline{\includegraphics[width=0.5\textwidth]{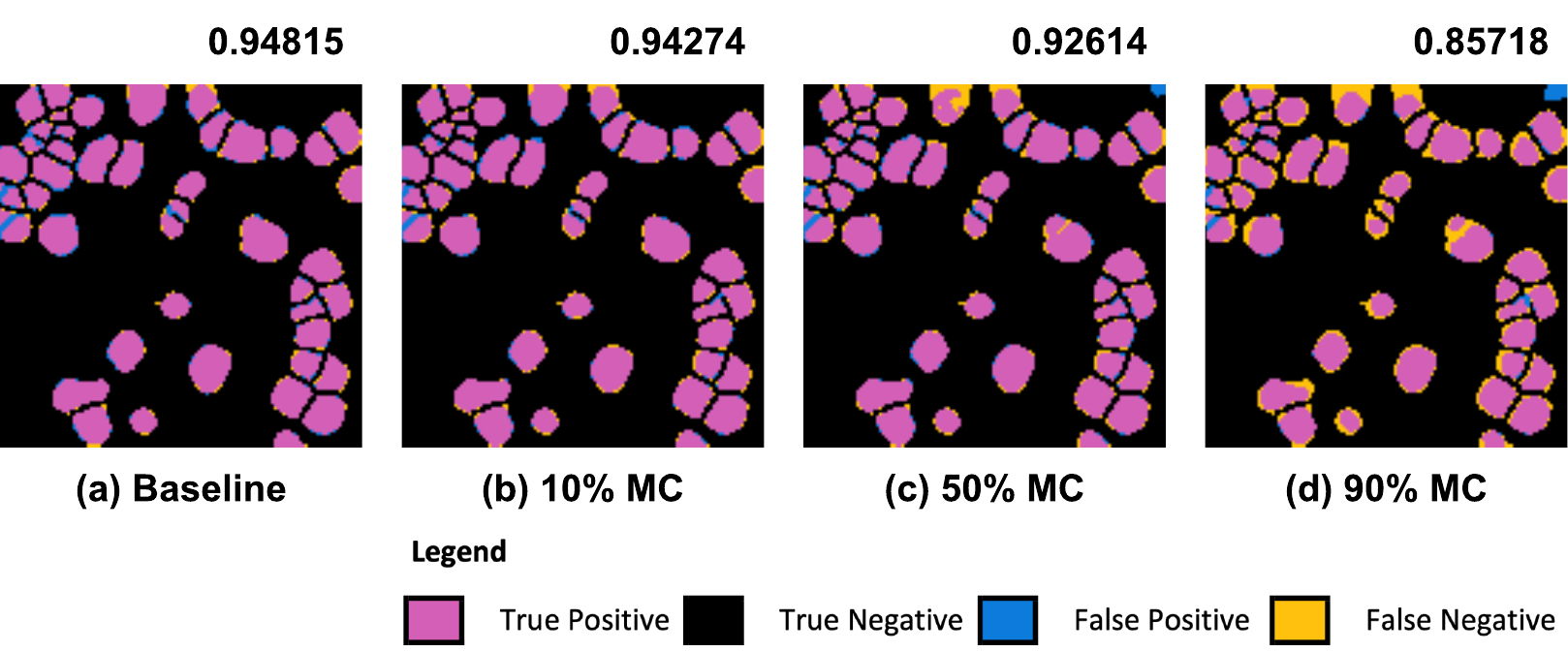}}
\caption{Example inference lung tissue patches with their corresponding \acrshort{dsc} scores for the (a) baseline, (b) 10\% \acrshort{mc}, (c) 50\% \acrshort{mc}, and (d) 90\% \acrshort{mc} models. With true positives in pink, true negatives in black, false positives in blue, and false negatives in yellow.}
\label{fig-patches-mc}
\end{figure}

The boxplot in Fig. \ref{fig-boxplot-mc} reports the \acrfull{dsc} for the label corruption baseline and \acrshort{mc} models, indicating that models trained on masks with up to 50\% \acrshort{mc} were on par with the baseline (0.874 vs 0.889), and started to decline in segmentation performance from 75\% \acrshort{mc} onwards (0.845 to 0.720). For visual purposes, we have provided four visualisations in Fig. \ref{fig-patches-mc} comparing the ground truth masks to the lung tissue inference masks for the baseline, 10\% \acrshort{mc}, 50\% \acrshort{mc}, and 90\% \acrshort{mc} models, outlining where the true/false positives and true/false negatives occur.

Quantitative results for the under- and over-segmentation models are shown in Table \ref{tab-results-uoseg}. We noticed that the baseline model performed best for \acrshort{dsc} score, Jaccard, precision, and specificity, but the 7$\times$7 under- and over-segmentation model resulted in the best recall value. There was a steep decline in \acrshort{dsc} score, Jaccard, precision, and specificity from the 5$\times$5 under- and over-segmentation model onwards, whilst the 3$\times$3 model stayed mostly in line with that of the baseline.

\begin{table}
\caption{Performance of Models Trained With Under- and Over-segmented \acrshort{gt} Masks}
\begin{center}
\renewcommand{\arraystretch}{1.4}
\setlength\tabcolsep{4pt}
\begin{tabular}{|c||c|c|c|c|c|}
\hline
\textbf{\begin{tabular}[c]{@{}c@{}}Training\\ Configuration\end{tabular}} & \textbf{\acrshort{dsc}} & \textbf{Jaccard} & \textbf{Precision} & \textbf{Recall} & \textbf{Specificity} \\
\hline
\hline
\textbf{\begin{tabular}[c]{@{}c@{}}Baseline\\ (50\% MC)\end{tabular}} & \textbf{0.874} & \textbf{0.753} & \textbf{0.875} & 0.813 & \textbf{0.972} \\
\hline
\textbf{3$\times$3} & 0.861 & 0.733 & 0.854 & 0.809 & 0.962 \\
\hline
\textbf{5$\times$5} & 0.832 & 0.690 & 0.753 & 0.860 & 0.906 \\
\hline
\textbf{7$\times$7} & 0.831 & 0.686 & 0.731 & \textbf{0.887} & 0.885 \\
\hline
\textbf{9$\times$9} & 0.808 & 0.657 & 0.731 & 0.839 & 0.894 \\
\hline
\end{tabular}
\label{tab-results-uoseg}
\end{center}
\end{table}

\subsection{Tissue Type Transferability}\label{tissue-transfer}
Table \ref{tab-results-tissuetrans} shows the difference in performance on the breast tissue test dataset across two groups of tissue type transferability models: (i) models trained on lung tissue only, and (ii) models trained on lung and breast tissue training sets. Overall, we found that the 50\% \acrshort{mc} model had a minimal difference in \acrshort{dsc} score compared to that of the baseline model (-0.031 vs -0.027). The same can also be said for Jaccard (-0.041 vs -0.040), precision (-0.039 vs -0.046), and specificity (-0.029 vs -0.034).

\begin{figure}
\centerline{\includegraphics[width=0.5\textwidth]{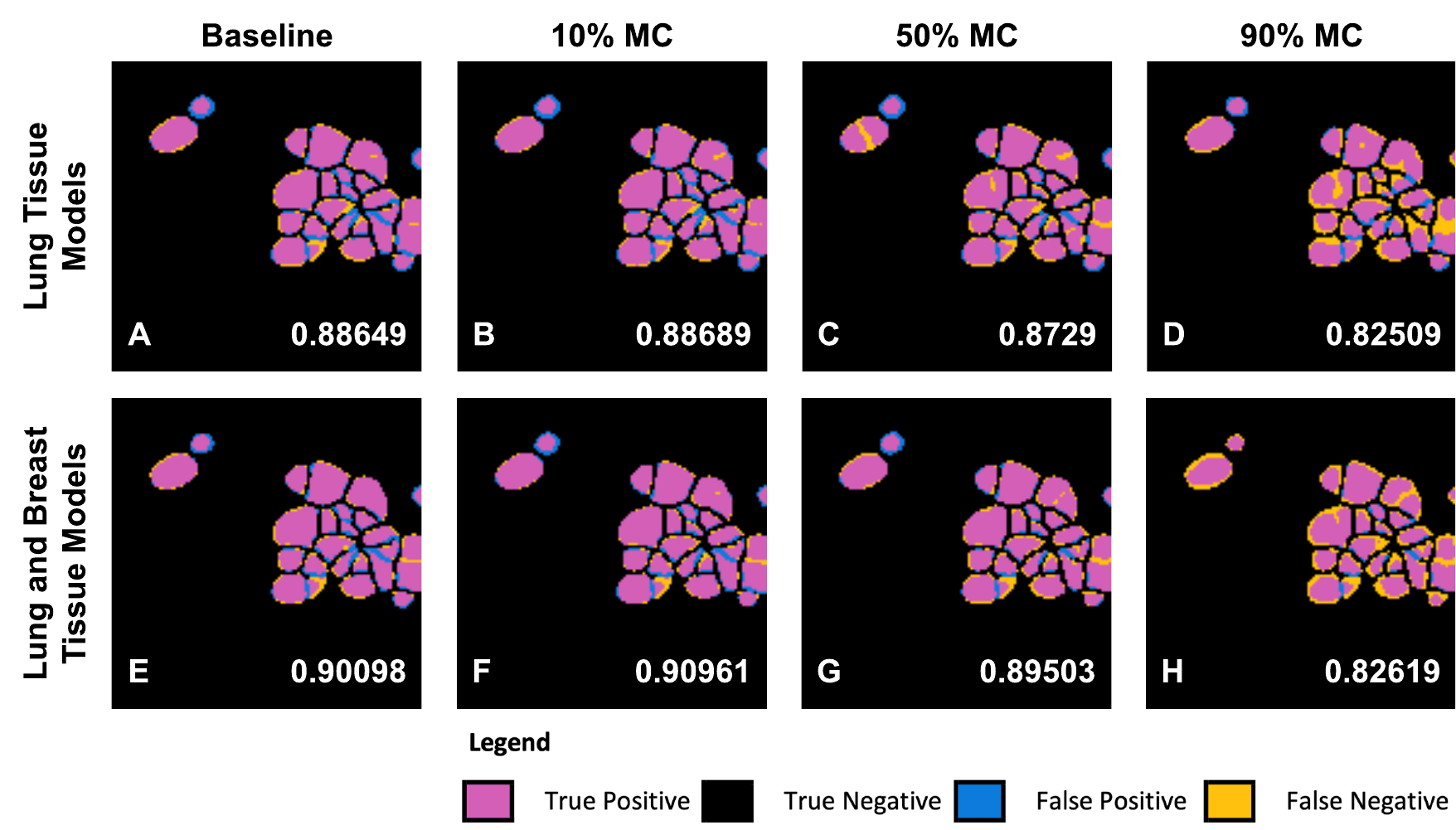}}
\caption{Example inference breast tissue patches with their corresponding \acrshort{dsc} scores for the baseline (A+E), 10\% \acrshort{mc} (B+F), 50\% \acrshort{mc} (C+G), and 90\% \acrshort{mc} (D+H) models trained on the lung tissue dataset and trained on both the lung and breast tissue datasets. With true positives in pink, true negatives in black, false positives in blue, and false negatives in yellow.}
\label{fig-patches-tissuetrans}
\end{figure}

Fig. \ref{fig-patches-tissuetrans} shows eight visualisations comparing the ground truth masks to the breast tissue inference masks for the baseline, 10\% \acrshort{mc}, 50\% \acrshort{mc}, and 90\% \acrshort{mc} models, outlining where the true/false positives and true/false negatives occur. The first four models (Figs. \ref{fig-patches-tissuetrans}A--D) were trained using only the lung tissue dataset, and the latter four models (Figs. \ref{fig-patches-tissuetrans}E--H) were trained using both the lung and breast tissue datasets.

\begin{table}[htbp]
\begin{center}
\renewcommand{\arraystretch}{1.4}
\setlength\tabcolsep{1.5pt}
\caption{Tissue Type Transferability Results. $\Delta M = M_{Lung} - M_{LungBreast}$, For Each Metric M}
\begin{tabular}{|c||c|c|c|c|c|}
\hline
\textbf{\begin{tabular}[c]{@{}c@{}}Training\\ Configuration\end{tabular}} & \textbf{$\Delta$\acrshort{dsc}} & \textbf{$\Delta$Jaccard} & \textbf{$\Delta$Precision} & \textbf{$\Delta$Recall} & \textbf{$\Delta$Specificity} \\
\hline
\hline
\textbf{\begin{tabular}[c]{@{}c@{}}Baseline\\ (0\% MC)\end{tabular}} & \textbf{-0.027} & \textbf{-0.040} & -0.046 & -0.002 & -0.034 \\
\hline
\textbf{10\% MC} & -0.032 & -0.048 & -0.055 & 0.008 & -0.057 \\
\hline
\textbf{50\% MC} & -0.031 & -0.041 & \textbf{-0.039} & -0.019 & \textbf{-0.029} \\
\hline
\textbf{90\% MC} & -0.038 & -0.048 & -0.132 & \textbf{0.059} & -0.071 \\
\hline
\end{tabular}
\label{tab-results-tissuetrans}
\end{center}
\end{table}

\subsection{Bootstrapping}\label{bootstrap}
The bootstrapping experiments were run on the 95\% \acrshort{mc} lung tissue model a total of ten times. We saw a total increase of 0.108 \acrshort{dsc} score from the baseline model up to the final bootstrapping iteration (\acrshort{dsc} 0.829), with a maximum \acrshort{dsc} score increase between two iterations of 0.028.

\section{Discussion}\label{discussion}
Throughout this work, we saw evidence suggesting that automated segmentation may not need as precise annotations as previously thought. In Fig. \ref{fig-boxplot-mc} we observe an approximately monotonic performance drop as a function of missing cells, with \acrshort{dsc} remaining stable up until the 50\% \acrshort{mc} model. It is only from the 75\% \acrshort{mc} model that we see a faster decline in \acrshort{dsc}, with the steepest decline being between the 90\% and 95\% \acrshort{mc} models. Fig. \ref{fig-patches-mc} also evidences insignificant changes in segmentation performance between models trained on labels with few cells missing as opposed to most cells missing. It shows a slight increase in false negatives in the 50\% \acrshort{mc} model (Fig. \ref{fig-patches-mc}c) and somewhat more in the 90\% \acrshort{mc} model (Fig. \ref{fig-patches-mc}d); this increase is primarily on cell edges and very small sections within cells. However, this is generally expected given just how many cell labels are missing from the \acrshort{gt} masks. While small amounts of under- and over-segmentation have a relatively low impact on performance, more detrimental effects are observed as kernel sizes increase (Table \ref{tab-results-uoseg}). However, we note that the recall improves in the 7$\times$7 under- and over-segmentation model (from 0.813 in the baseline model to 0.887). There is also a much larger drop in \acrshort{dsc} when going from 3$\times$3 to 5$\times$5 under- and over-segmentation (a difference of 0.029 \acrshort{dsc} score). It is clear from this that the segmentation models are not as robust to under- and over-segmentation compared to missing cells. 

The two groups of tissue type transferability models produce minimal differences in \acrshort{dsc} scores for the baseline, 10\% \acrshort{mc}, and 50\% \acrshort{mc} models (-0.027, -0.032, and -0.031 respectively), of which is shown in Table \ref{tab-results-tissuetrans}. There is a slightly larger \acrshort{dsc} score difference for the 90\% \acrshort{mc} models (-0.038), however, this is still not very significant. Generally, all four models have a better performance when training on both the lung and breast tissue datasets, however, models trained solely on the lung dataset still perform very similarly and can still be considered usable. Fig. \ref{fig-patches-tissuetrans} shows very few negative effects by training solely on the lung tissue dataset, for example, for both the baseline and 10\% \acrshort{mc} models there are only slightly more false negatives in the lung tissue model patches compared to the lung and breast tissue model patches. This is also the case for the 50\% \acrshort{mc} and 90\% \acrshort{mc} models but appears only to affect a small number of cell edges. Moreover, the results of the bootstrapping experiments imply that models with fewer training labels have the ability to improve themselves using the results of the previous iteration's inference, suggesting that models may not necessarily be impacted by their smaller set of \acrshort{gt} masks.\\
Overall, the results presented suggest that imprecise and inaccurate \acrshort{gt} labels can confer accurate \acrshort{imc} cell segmentation results in deep learning contexts. The results of label corruption experiments indicate that annotators can save at least half of their time labelling \acrshort{imc} images for training purposes. Furthermore, tissue-specific segmentation models may not be necessary in contexts where there are corresponding \acrshort{imc} biomarker channels available.

\section{Conclusion}\label{conclusion}
This work studies the effects that varying levels of label imperfections (including missing cells, and under- and over-segmentation) in \acrshort{imc} data can have on learning-based segmentation models, by utilising classical image processing approaches to simulate imperfections. Additionally, we study the generalisability of these models to different \acrshort{imc} tissue types and investigate if bootstrapping models with significantly fewer cell labels is a viable option. Results suggest that training models on around half of the number of labels as provided in original \acrshort{gt} masks can offer segmentation results on par with models provided with complete annotations; implying that current workload annotating \acrshort{imc} images can in fact be cut down by at least half. On top of that, findings from tissue transferability experiments suggest that future models may not necessarily need to be trained on varying tissue types, and are transferable to one another. Furthermore, models with significantly fewer training labels can improve themselves using their own inference masks.

\section{Compliance with Ethical Standards}\label{ethics}
This research study was conducted retrospectively using human subject data made available in open access by \cite{rendeiro2021} and \cite{ali2020}. Ethical approval was not required as confirmed by the license attached with the open access data.


\begin{thebibliography}{00}
\bibitem{giesen2014} C. Giesen et al., ``Highly multiplexed imaging of tumor tissues with subcellular resolution by mass cytometry,'' Nature Methods, vol. 11, pp. 417--422, 2014.
\bibitem{glasson2023} Y. Glasson et al., ``Single-cell high-dimensional imaging mass cytometry: one step beyond in oncology,'' Semin Immunopathol, vol. 45, pp. 17--28, 2023.
\bibitem{melin2022} N. Melin et al., ``A new mouse model of radiation-induced liver disease reveals mitochondrial dysfunction as an underlying fibrotic stimulus,'' JHEP reports, vol. 4, 2022.
\bibitem{zabransky2023} D. J. Zabransky et al., ``Profiling of syngeneic mouse HCC tumor models as a framework to understand anti-PD-1 sensitive tumor microenvironments,'' Hepatology, vol. 77, pp. 1461--1462, 2023.
\bibitem{greenwald2022} N. F. Greenwald et al., ``Whole-cell segmentation of tissue images with human-level performance using large-scale data annotation and deep learning,'' Nature Biotechnology, vol. 40, pp. 555--565, 2022.
\bibitem{mund2022} A. Mund et al., ``Deep visual proteomics defines single-cell identity and heterogeneity,'' Nature Biotechnology, vol. 40, pp. 1231--1240, 2022.
\bibitem{baars2021} M. J. D. Baars et al., ``MATISSE: a method for improved single cell segmentation in imaging mass cytometry,'' BMC Biology, vol. 19, 2021.
\bibitem{ronneberger2015} O. Ronneberger, P. Fischer, and T. Brox, ``U-Net: convolutional networks for biomedical image segmentation,'' Lecture Notes in Computer Science, vol. 9351, pp. 234--241, October 2015 [MICCAI, p.3, 2015].
\bibitem{rendeiro2021} A. F. Rendeiro et al., ``The spatial landscape of lung pathology during COVID-19 progression,'' Nature, vol. 593, pp. 564--569, 2021.
\bibitem{ali2020} H. R. Ali et al., ``Imaging mass cytometry and multiplatform genomics define the phenogenomic landscape of breast cancer,'' Nature Cancer, vol. 1, pp. 163--175, 2020.
\bibitem{berg2019} S. Berg et al., ``ilastik: interactive machine learning for (bio)image analysis,'' Nature Methods, vol. 16, pp. 1226--1232, 2019.
\bibitem{stirling2021} D. R. Stirling et al., ``CellProfiler 4: improvements in speed, utility and usability,'' BMC Bioinformatics, vol. 22:433, 2021.
\end{thebibliography}
\end{document}